# WHY ARE MEGAPROJECTS, INCLUDING NUCLEAR POWER PLANTS, DELIVERED OVERBUDGET AND LATE? REASONS AND REMEDIES

Dr Giorgio Locatelli

University of Leeds

g.locatelli@leeds.ac.uk

# 1 Introduction to the realm of infrastructure megaprojects

## 1.1 The Performance of Megaprojects

Megaprojects are usually defined as projects with budgets above $1 billion that involve a high level of innovation and complexity (Flyvbjerg et al. 2003; Locatelli, Mancini, et al. 2014; Merrow 2011; Van Wee 2007). However, already in 1985, (Warrack 1985) noted that $1 billion is not a constraint in defining megaprojects since sometimes a relative approach is needed. In fact, in some contexts, a much smaller project (such as one with a $100 million budget), could constitute a megaproject. Similarly, (Hu et al. 2013) claim that a deterministic cost threshold is not appropriate for all countries, and a relative threshold such as the GDP should be used instead. According to (Brookes & Locatelli 2015) *"Megaprojects are temporary endeavours (i.e. projects) characterized by large investment commitment, vast complexity (especially in organizational terms), and long-lasting impact on the economy, the environment, and society"*. Megaprojects tend to be massive, indivisible, and have long-term impacts, with investments taking place in waves. Megaprojects' *"effects are felt over many years, especially as auxiliary and complementary additions are made"* (Miller & Lessard 2001). Public policy strongly affects the performance of public megaprojects. In fact, megaprojects *"remain under political scrutiny well after the final official decision is made. Decisions made early on can have disastrous effects when abstract political ambitions crystalize in specific technical challenges"* (Giezen 2012). The planning and delivery of megaprojects involves a large number of organisations and sometimes planning and delivery last longer than the organisations themselves (Brookes et al. 2017). Clearly, Nuclear Power Plants (NPP) fit the definition and characteristics of megaprojects.

Despite their fundamental economic and social role, megaprojects are extremely risky (Locatelli & Mancini 2010) and are often implemented after a weak (i.e. often sub-optimal) phase of project planning leading to underestimations of the costs and overestimation of short-term benefits (Flyvbjerg 2006). An analysis of data from 318 industrial megaprojects (Merrow 2011) shows that the vast majority of megaprojects might be considered a failure when considering adherence to schedule and budget as well as benefits in operation. Megaprojects are often affected by budget overruns, delays in different phases of the project development, and their operating results rarely justify, in the short term at least, the implementation of the project (Flyvbjerg et al. 2003)[1].

Table 1 provides an overview of large and mega transportation projects, demonstrating how these projects are overbudget all over the world. The cost escalation is common even for long-planned, multi-billion dollar, one-of-a-kind infrastructure projects such as Eurotunnel which increased by 59% alone or 69% considering other related projects (Winch 2013).

---

[1] It is worth noting that authors such as (Dimitriou et al. 2013; Turner & Zolin 2012) are quite critical of Flyvbjerg and Merrow's perspective about the benefits of megaprojects, and stress that they need to be assessed with a much longer-term perspective and include a wide range of stakeholders.

| Reference | Location | Sector/Infrastructure | Sample Size | Average Over budget |
|---|---|---|---|---|
| (Flyvbjerg et al. 2016) | World | Roads | 863 | + 20% |
| | Hong Kong's | Roads | 25 | +11% / +6% / -1 % |
| (Cantarelli & Flyvbjerg 2015) updating (Flyvbjerg 2008) | World | Rail | 58 | +45% |
| | | Fixed Link | 33 | + 34% |
| | | Road | 167 | + 20% |
| (Cantarelli & Flyvbjerg 2015) | Europe | Rail | 23 | +34% |
| | | Fixed Links | 15 | +43% |
| | | Road | 143 | +22% |
| | | Total | 181 | +26% |
| | North America | Rail | 19 | +41% |
| | | Fixed Links | 18 | + 26% |
| | | Road | 24 | +8% |
| | | Total | 61 | +24% |
| | Other geographical areas | Rail | 16 | +65% |
| (Cantarelli et al. 2012) | Netherlands | Road | 37 | +19% |
| | | Rail | 26 | + 11% |
| | | Fixed Links | 15 | + 22% |
| (Lee 2008) cited in (Cantarelli et al. 2012) | South Korea | Road | 138 | + 11% |
| | | Rail | 16 | +48% |

**Table 1 Review of Large Transportation projects studies. Extract from (G. Locatelli et al. 2017)**

## 1.2 Reasons for cost escalation/overbudget

The question of megaprojects often being overbudget and late has been addressed by many researchers in the field. Three stand out as major contributors: Bent Flyvbjerg, Edward Merrow and Peter E.D. Love. Their views are summarized in this section.

### 1.2.1 Perspective of Bent Flyvbjerg

Bent Flyvbjerg is the Professor and inaugural Chair of Major Programme Management at Oxford University's Saïd Business School. (Cantarelli et al. 2010) summarizes the most important arguments that have emerged from Flyvbjerg's studies, e.g. (Flyvbjerg 2006; Flyvbjerg & Molloy 2011; Flyvbjerg et al. 2016) as follows:

- **Technical:** Forecasting errors including price increases, poor project design, and incomplete estimates, scope changes, uncertainty, inappropriate organizational structure etc.
- **Psychological:** optimism bias among local officials, the cognitive bias of people, and cautious attitudes towards risk.
- **Deliberate underestimation of costs from:**
  - **Vendors/contractors:** willingness of vendors to intentionally underestimate the cost and overestimate the benefit in order to "sell the projects".

- - **Political:** manipulation of forecasts for starting projects for unethical reasons (e.g. gain consensus for the next elections) or corruption phenomena in general (see also (G. Locatelli et al. 2017) on this latter point).
- **Poor financing / contract management:** poor risk allocation, typically awarding the project (or sub-contracts) the lowest priced "turn-key contract" allocating the key risks to contractors that might not be able to properly manage them (see for instance the case of concrete in Olikuoto 3 - (Ruuska et al. 2009)).

Deliberate underestimations and psychological reasons are key in Flyvbjerg's research. However, there is an important distinction to make, as clearly presented in (Flyvbjerg 2006) *"Psychological and political explanations better account for inaccurate forecasts. Psychological explanations account for inaccuracy in terms of optimism bias, that is, a cognitive predisposition found with most people to judge future events in a more positive light than is warranted by actual experience. Political explanations, on the other hand, explain inaccuracy in terms of strategic misrepresentation. Here, when forecasting the outcomes of projects, forecasters and managers deliberately and strategically overestimate benefits and underestimate costs in order to increase the likelihood that it is their projects, and not the competition's, that gain approval and funding. Strategic misrepresentation can be traced to political and organizational pressures, for instance, competition for scarce funds or jockeying for position. Optimism bias and strategic misrepresentation are both deceptions, but where the latter is intentional, i.e., lying, the first is not, optimism bias is self-deception. Although the two types of explanation are different, the result is the same: inaccurate forecasts and inflated benefit-cost ratios".*

So according to Bent Flyvbjerg the specific characteristics of each megaproject are rather irrelevant, and the overbudget /delay is explained by the stakeholder's attitude toward the project.

### 1.2.2 Perspective of Edward Merrow

Edward Merrow is a leading practitioner, founder and President of Independent Project Analysis, Inc. (IPA) and has spent over 35 years studying megaprojects at IPA and the Rand Corporation. According to Merrow (2011) the top seven mistakes leading to overbudget and delay in megaprojects are:

1. Greed
2. Schedule pressure to reduce construction time and increase the Net Present Value
3. Poor bidding phase

4. Reduction in the upfront cost, leading to poor quality Front End Loading (FEL) and Front End Engineering and Design (FEED)[2].
5. Unrealistic cost estimations
6. Poor risk allocation
7. Excessive pressure on project manager and "blame culture"

Merrow (2011) emphasizes the importance of a very well developed FEL affirming that " *After 30 years of showing the data, badgering, cajoling, and whining to the industry about the criticality of FEL, I believe there is now virtual consensus among project professionals within the community of industries we serve that FEL is the single most important predictive indicator of project success*".

Figure 1 shows the relationship between the quality of the FEL and the cost performance. There is a very clear trend showing how FEL is a major determinant of the project performance. If the FEL is properly developed, the budget is usually adhered to; if the FEL is poor or is just at a concept phase, cost deviation of 50% or more can be expected. A similar discussion can be made for the schedule (i.e. the time performance). (Merrow 2011) assesses the quality of the FEL by examining its three phases. Phase is 1 is about assessing the "business case", "team dynamics", and "alternatives analysis", each of these elements with its own metrics (e.g. market experience). Phases 2 and 3 are about assessing "site factors", "design status", "project execution plan", each of these elements with its own metrics, more generic in phase 2 (e.g. availability of preliminary soils and hydrology report) and more specific in phase 3 (e.g. quality of the information about labor availability, cost and productivity).

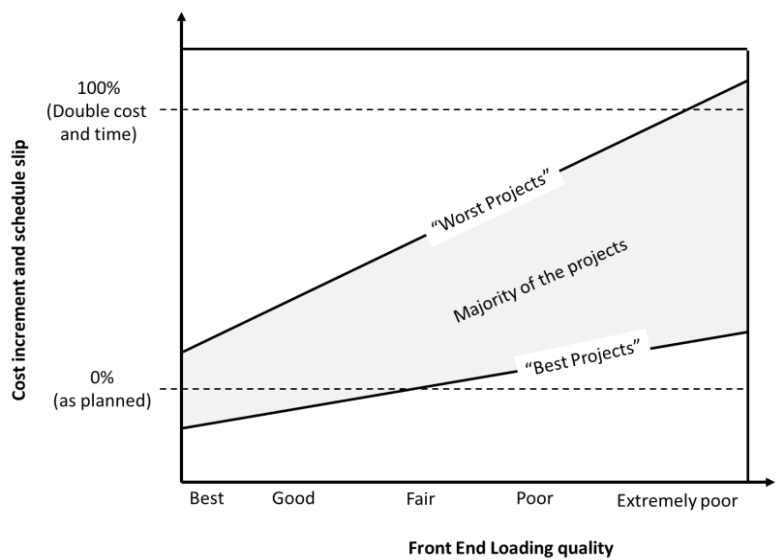

**Figure 1 Effect of Front End Loading on Project Management performance. Adapted from (Merrow 2011)**

---

[2] Front-end loading (FEL) is generally defined as the process starting from the appraisal of the project opportunity, and leading to the development of the project scope and ultimately to the project definition. Project definition includes the development of the final feasibility study and front-end engineering design (FEED), i.e. the conceptual design of the plant.

### 1.2.3 Perspective of Peter ED Love

Peter E.D. Love is a John Curtin Distinguished Professor at Curtin University. In his books and papers he provides an original perspective on the studies of megaprojects. (Love et al. 2016; Ahiaga-Dagbui & Smith 2014) recognize that two predominant schools of thoughts have emerged from the on-going discourse on cost overruns: (a) "evolution theorists" who state that overruns are caused by scope changes during the development of a project and (b) "psycho strategists" (championed by B. Flyvbjerg) who combine psychological contributors and business strategy (such as deception, planning fallacy and unjustifiable optimism) to explain the causes of cost overruns. In support of the "psycho strategist" point of view (Love et al. 2012) suggests that cost overruns arise as a *"result of pathogenic influences, which lay dormant within the project system"*.

However, (Love et al. 2016) also stress that optimism bias and strategic misrepresentations alone cannot adequately explain why transportation infrastructure projects suffer cost overruns, and the evidence presented by the psycho strategists lacks credibility. Therefore, (Love et al. 2015) have advocated a more balanced approach that focuses on both processes and technology innovations. Notably, (Love et al. 2016) point out the tendency to study independent causes of cost overruns, and the lack of investigation of the interdependencies among those causes. To reinforce this argument, Love (Love et al. 2016) uses the two and three variable path model as shown in Figure 2 to address the "Simpson's paradox".

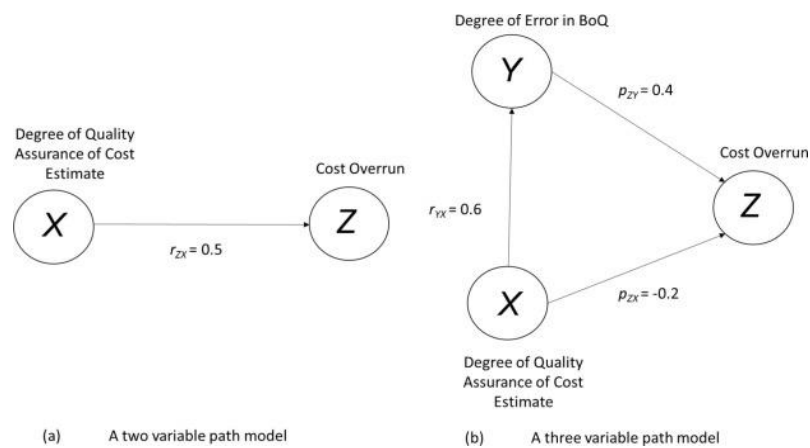

**Figure 2 Path model demonstrating Simpson's Paradox. Picture from** (Love et al. 2016)**. BoQ = Bill of Quantities; $r_{zx}$ is the "apparent" correlation between "X" and "Z"; $p_{zx}$ is the "real" correlation between "X" and "Z" when the mediating element Y is considered; $r_{yx}$ is the "real" correlation between "X" and "Y"; $p_{zy}$ is the "real" correlation between "Y" and "Z"**

Let us take the example of a tall professional model that is used to walk with high heels every night, in any weather condition. If we investigate independently what causes her to fall on a rainy night, we might select the following independent variables: (1) she is a tall professional model, (2) she wears high heels, (3) it rained on the night she fell.

Testing variable (1) independently will make us assume that the fact that she is a tall professional model is something that might cause her to fall (part (a) in Figure 2). However, reasonably, it might be that it is the combination of variables (1) and (2) (or (1) and (3), or (1), (2) and (3) together) that is correlated to the probability of falling (part (b) in Figure 2).

Considering the nuclear sector, the correlation "Nuclear project" with "Overbudget and delay" might be investigated looking at the possible "mediating factors". E.g. if the mediating factor is "lack standardization" (intended both as physical and supply chain, see section 2) the efforts should be focused on reducing this "lack of standardization". In other words, there is nothing wrong with nuclear projects "per se", but the problem lies on how are planned and delivered.

Therefore Love is more circumspect in explaining the overbudget by people behavior, and it is more focused on the scope and content of the project itself. In a nutshell according to Love strategic and economic decisions taken for a project influences the way in which an organization processes information, which affects the way they manage risk. In the nuclear sector, this relates to what is explained in sections 2.

### 1.2.4 Other perspectives

Another reason for megaprojects overbudget is the so-called "winner curse". The winner's curse is *"a tendency for the winning bid in an auction to exceed the intrinsic value of the item purchased. Because of incomplete information, emotions or any other number of factors regarding the item being auctioned, bidders can have a difficult time determining the item's intrinsic value. As a result, the largest overestimation of an item's value ends up winning the auction"* (Investopedia 2017). So, in simple terms, the infrastructure that appears to have the most optimistic cost/benefit analysis is usually selected. This infrastructure is the "most optimistic", therefore the one more likely to be overbudget/late.

The problem of budget overruns in megaprojects is systematic with no relevant improvement over time. This seems to be an example of "*tolerance for deviation*" (Pinto 2014), i.e. people within the organizations become so accustomed to a deviant behaviour that they do not consider it as deviant anymore. Normalization of deviance suggests that the unexpected becomes the expected[3]. In particular, politicians play the necessary "political games" and maintain important contacts to ensure broad-based support for the project despite the poor project management performance (Pinto &

---
[3] This concepts applies very well to the nuclear industry in USA and and Europe. Are we really surprised that all those plants, none excluded, are overbudget and late?

Patanakul 2015). Locatelli's research (Brookes & Locatelli 2015; Giorgio Locatelli, Mikić, et al. 2017; Locatelli, Invernizzi, et al. 2017; Wang et al. 2017; Leurent et al. 2017) also links a series of project characteristics to successful performance in terms of avoidance of delays and cost overruns. The project environment, and its legal and socioeconomic characteristics in particular have been identified as having an important relationship with megaprojects success. For example, the influence of Special Purpose Entities[4] (Sainati et al. 2017) is of particular importance for the project management success. Recently an empirical analysis of 44 megaprojects (Locatelli, Mikić, et al. 2017) (including 5 nuclear projects) delivered in Europe concluded that the successful delivery of megaprojects to be secured, projects need to (a) engage better with external stakeholders of the megaproject (and especially environmental groups), the affected population and regulators and (b) understand how to make the best use of Special Purpose Entities in the governance of megaprojects.

## 1.3  Improving the quality of cost estimation: the reference class forecast

A commonly suggested solution to cure the poor budget forecast (particularly considering optimism bias and strategic misinterpretation) is the so-called "reference class forecast" as presented in (Flyvbjerg 2006). Reference class forecasting *"is a method for systematically taking an outside view on planned actions. More specifically, reference class forecasting for a particular project requires the following three steps:*

*(1) Identification of a relevant reference class of past, similar projects. The class must be broad enough to be statistically meaningful but narrow enough to be truly comparable with the specific project.*
*(2) Establishing a probability distribution for the selected reference class. This requires access to credible, empirical data for a sufficient number of projects within the reference class to make statistically meaningful conclusions.*
*(3) Comparing the specific project with the reference class distribution, in order to establish the most likely outcome for the specific project.*

*Thus reference class forecasting does not try to forecast the specific uncertain events that will affect the particular project, but instead places the project in a statistical distribution of outcomes from the class of reference projects. In statisticians vernacular, reference class forecasting consists of regressing forecasters' best guess toward the average of the reference class and expanding their estimate of credible interval toward the corresponding interval for the class".*

---

[4] According to (Sainati et al. 2017) a Special Purpose Entity is a fenced organization having limited predefined purposes and a legal personality. Special Purpose Entities are often used to design, deliver, finance, and operate infrastructural mega-projects. Their relevance lies in the ability to attract finances, manage the risks, and shape the governance of the megaproject. Corporate Partnership/Joint Venture, Project Joint Venture, Public Private Partnership etc. are usually based on Special Purpose Entities.

It is important to stress that the reference class forecast does not aim to reduce the cost of a project but to limit the project's potential for being overbudget by increasing the quality of (and confidence in) the cost estimation. Nevertheless, the reference class forecast presents also some limitations during its actual application, especially in the case where projects are technically, economically, politically and socially complex and very different from one another (e.g. the case of new NPPs with different designs from the past), and/or the number of comparable projects is too low (e.g. the case of nuclear decommissioning projects). If the projects are not radically different, a modified version of the reference class forecast could be done using construction time instead of budget, as mentioned in section 2.2. Time comparison is more reliable since monetary values need to be adjusted for inflation, financing terms, engineering cost, currency exchange etc. Construction time is a more immediate measure of performance. If the NPP design/projects are radically different, the reference class forecast method is not applicable. In these cases, benchmarking analysis could be adapted to the specific projects characteristics to evaluate their impact on the actual project performance, see for example (Invernizzi et al. 2017a; Invernizzi et al. 2017b; Locatelli, Invernizzi, et al. 2017)

# 2 Nuclear power plants as megaprojects

The previous section reviewed the general topic of megaprojects, with their project management performance, their causes and the role of the reference class forecast. These arguments apply to NPPs as well. However NPPs, given their size and complexity, are a special type of megaprojects, so further considerations are needed.

## 2.1 Overbudget and delays in the nuclear sector

### 2.1.1 The role of the project size

NPPs are among the largest type of infrastructure that have been built. "Largest" here is intended as physically massive, with a budget in the range of tens of billions of dollars, and delivery (planning + construction) taking up to decades. Examples include:

- EPR Olkiluoto 3, with a timeline of about 18 years (2000 - TVO application Finnish cabinet for a decision-in-principle; operation expected in 2019 (WNN 2017b), final cost unknown but probably close to €8.5 billion (Milne 2016);

- EPR Flamanville 3, with a timeline of about 13 years (October 2005, beginning of the national public debate, operation expected in 2019 (WNN 2017a)), a final budget above €10 billion (EDF 2015);

- AP1000: costs to build the two Vogtle units could range between $18.3 billion and $19.8 billion (Patel 2017). The AP1000 units in Virgil C. Summer have comparable budget, and the projects are currently (September 2017) suspended ;

So, surely the construction of a GWe scale NPP can be considered as a megaproject, and surely one of the "largest megaproject". The key issue is that the larger the project, the greater is the likelihood to be over budget and experience a delay. Indeed, this fact is supported by several empirical studies as explained next, and it is a key determinant for the performance of nuclear projects.

Regarding the correlation of size with overbudget, one of the most holistic and recent studies (Ansar et al. 2014) analyzed the budget and schedule performance of 245 large dams, built between 1934 and 2007, on five continents, in 65 different countries. It was found that 3 out of 4 large dams suffered cost overruns and delays. Actual costs were on average 96% higher than estimated costs; the median was 27%. This large difference between the average and median suggests that a few megaprojects were enormously overbudget, but probably also that they were outliers, and a more "moderate" overbudget is common (see Table 1). Differences among regions are not significant. Project type (e.g., hydropower, irrigation, or multi-purpose dam) or wall type (earth fill, rock fill, concrete arch, etc.) did not influence the cost overruns. Irrespective of the year or decade in which a dam was built, there were no significant differences in forecasting errors, implying little learning from past mistakes. Quite remarkably, the only statistical significance is that the level of cost overruns increases with dam size measured either by installed hydropower generation or wall height. Similar results were found for the schedule.

The analysis of a dataset of 401 power plants and transmission projects in 57 countries, showed that only 39 projects across the entire sample experienced no cost overrun (Sovacool, Nugent, et al. 2014). Later, the same authors (Sovacool, Gilbert, et al. 2014a) have shown that technologies usually delivered as megaprojects are most likely to be over budget, in fact, *"Hydroelectric dams and nuclear reactors have the greatest amount and frequency of cost overruns, even when normalized to overrun per installed MW […] solar and wind projects seem to present the least construction risk."* The idea that the bigger the project, the more likely it was to have a budget overrun was found to be statistically significant even in (Sovacool, Gilbert, et al. 2014b). (Brookes & Locatelli 2015) analysed a sample of 12 energy megaprojects and investigated the project characteristics that correlated with the project performance. Again, the construction of nuclear NPPs (the largest type of project in the sample) strongly correlates with budget overruns and have the largest cost overrun.

### 2.1.2 The role of the project complexity

Project complexity (technical and organizational) is another major determinant of overbudget and delay in the nuclear sector. (Locatelli, Mancini, et al. 2014) investigated the role of complexity in megaprojects, highlighting that underperforming projects are often delivered in a project environment characterized by:

1. rapid changes of technologies; shortened technology cycle time; increased risks of obsolescence (Hanratty et al. 2002);

2. increasingly interoperable and interdependent systems (Jaafari 2003);

3. emphasis on cost reduction, with tight schedules and without quality or scope reduction (Laufer et al. 1996);

4. integration issue: high number of system parts and organizations involved (Calvano & John 2004)(Locatelli & Mancini 2010);

5. combining multiple technical disciplines (Ryan & Faulconbridge 2005);

6. competitive pressures from other technologies within the same market (CCGT vs nuclear) or other designs within the same technology (Kossiakoff et al. 2011).

These six elements are typical for technically complex projects, element 1 being the exception for nuclear. With this background and the guidelines of (GAPPS 2007), (Locatelli, Mancini, et al. 2014) further defined a metric they termed "complex project context", i.e. where the complexity is not only technical, but also organizational in the "project delivery chain". Specific characteristics include:

- several key distinct disciplines, methods, or approaches involved in performing the project;
- strong legal, social, or environmental implications from performing the project;
- usage of most of partner's resources (both tangible and intangible);
- the strategic importance of the project to the organization or organizations involved;
- stakeholders with conflicting needs regarding the characteristics of the product of the project; and
- high number and variety of interfaces between the project and other organizational entities.

All these aspects have major influence on the project governance (Müller 2009) since *"managerial rationalities are limited in understanding their own complex project realities which are themselves bound by limits imposed by overall governance structures and strategies"* (van Marrewijk et al. 2008). Therefore the planning and delivery of NPPs is surely complex, where complexity is intended to encompass both technical/physical and organizational aspects of the project.

The first layer of complexity is technical/physical complexity, as NPPs are characterized as big engineering challenges and involve a diversity of disciplines[5]. An NPP project includes nuclear engineering (that is mostly peculiar for this type project), but also, civil, electrical, I&C, mechanical, hydraulic, materials, etc. Outside engineering, there are challenges of physics, geology, human-machine interaction, cyber security etc.

The second layer of complexity is organisational. NPPs, by nature, are subject to very strict and unique regulations, that have many undesirable effects, e.g. increasing the complexity and cost of the supply chain. The organizational complexity also comes from the political involvement, since nuclear energy can be a very sensitive topic in the public discourse. NPPs require specific skills that need to be maintained and developed, e.g. through academia and training. Moreover, the financing itself is a major issue since the development of an NPP project is a very risky project with capital in the range of billions invested with no revenue generation for at least 5 years and more typically a decade. It is a challenge to find investors with such a large amount of financial resources and risk appetite.

---

[5] There are very few projects with comparable complexity. Examples might be the space missions or large fusion experiments, other sectors very well known for their Project Management issues (regarding fusion, see the recent case of the ITER experimental reactor)

## 2.2 Application of the reference class forecast to the nuclear sector

An application of the reference class forecast to the nuclear sector is presented in (Locatelli & Mancini 2012). In this paper, the authors investigated the case of Olkiluoto 3 and Flamanville 3 and benchmarked these NPPs against historical data.

In both cases, the projects have significantly underestimated the time and the resources required to complete construction. While the failure to reach construction targets may be due to FOAK (First of A Kind) effects that are hard to predict, failure to set targets may also have been due to the unjustified optimism in the process to reach the target. To gauge the realism of the targets used in Flamanville 3 and Olkiluoto 3, it is useful to apply reference forecasting in order to evaluate the gap between planned and actual cost and schedule data of the project. The construction of EPR was expected to take about 5 years for Flamanville 3 and Olkiluoto 3. (Grubler 2010) clearly shows that an increase in size increases the construction time, and the EPR is bigger than any other NPP. Moreover, the EPR is based on the German NPP "Konvoi" and the French NPP "N4". The decision to develop and build the N4 NPP was the most problematic of the entire French PWR (Pressurised Water Reactors) program: the new NPPs faced numerous technical difficulties, substantial delays, and by French standards prohibitive cost overruns. Not a single N4 NPP was exported. Focusing on these NPPs, Table 2 and Figure 3 show that the initial forecasts for the EPR projects were too optimistic when compared to previous values of NPPs built in France. The previous reference NPPs were completed in about 10 years, and they had been built during the "golden age" of NPP when the entire French project delivery chain was gaining more and more experience in the construction of a number of nuclear NPPs, and FOAK effects were therefore minimised. The new EPRs are bigger, more complex and built using inexperienced supply chains. Nevertheless, the initial forecast was a 50% reduction in the schedule: both forecasts for Flamanville 3 and Olkiluoto 3 demonstrated an optimism bias. It is clear that the actual values match the historical performance, as expected with a "reference class" forecast.

| State | NPPs Name | Type | Location | Net Capacity | Construction Started | Commercial Operation | TOTAL Const time [years] |
|---|---|---|---|---|---|---|---|
| France | CHOOZ-B-1 | N4 | ARDENNES | 1500 | 01/01/1984 | 15/05/2000 | 16.4 |
| France | CHOOZ-B-2 | N4 | ARDENNES | 1500 | 31/12/1985 | 29/09/2000 | 14.8 |
| France | CIVAUX-1 | N4 | VIENNE | 1495 | 15/10/1988 | 29/01/2002 | 13.3 |
| France | CIVAUX-2 | N4 | VIENNE | 1495 | 01/04/1991 | 23/04/2002 | 11.1 |
| Germany | BROKDORF (KBR) | Konvoi | BROKDORF | 1410 | 01/01/1976 | 22/12/1986 | 11.0 |
| Germany | PHILIPPSBURG-2 (KKP 2) | Konvoi | PHILIPPSBURG | 1402 | 07/07/1977 | 18/04/1985 | 7.8 |
| Germany | ISAR-2 (KKI 2) | Konvoi | ISAR | 1410 | 15/09/1982 | 09/04/1988 | 5.6 |
| Finland | Olkiluoto 3 | EPR | Olkiluoto | 1650 | 12/08/2005 | 15/05/2019 | 13.8 |
| France | Flamanville 3 | EPR | Flamanville | 1650 | 03/12/2007 | 01/11/2019 | 11.9 |

**Table 2 Construction time for EPR NPPs in Europe. Historical data from IAEA PRIS database www.iaea.org/pris/ forecast (WNN 2017b) for Olkiluoto 3 and (WNN 2017a) for Flamanville 3**

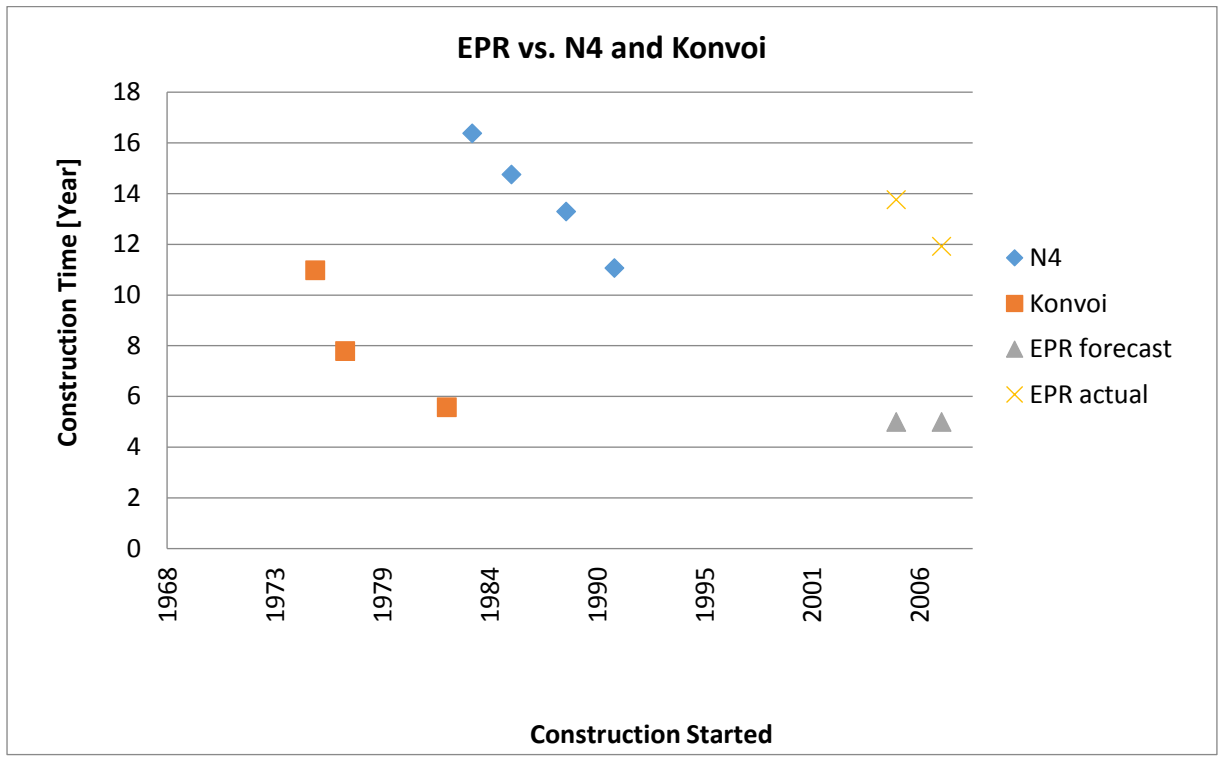

**Figure 3 Application of reference cast forecast. Updated from (Locatelli & Mancini 2012) using (WNN 2017b) for Olkiluoto 3 and** (WNN 2017a) **for Flamanville 3**

## 2.3 Potential approaches for cost reduction

### 2.3.1 Economy of multiples

The economy of multiples is achieved when the same identical plant (e.g. an NPP) is delivered more than one time, and ideally several times. Economy of multiples in the construction of NPP is somehow rooted in the idea of "mass production", a concept born in the automotive industry and later adopted in other fields, like aerospace (e.g. the production of aircraft), IT (e.g. the production of computers and smartphones) or even the food industry (e.g. the production of ready meals). For NPP, the economy of multiples is achieved because of two key factors: the learning process and the co-siting economy.

#### 2.3.1.1 Learning process

Learning economies result from the replicated supply of plant components by the suppliers and from the replicated construction and operation of the plant by the utilities and their contractors. Therefore, the unit cost of an Nth-of-A-Kind (NOAK) plant is less than a First-of-A-Kind (FOAK). The learning process reduces the cost of equipment, material and work. However, another result from learning is the reduction in the construction schedule. As shown in (Locatelli, Bingham, et al. 2014), the construction schedule is another very critical economic aspect in an NPP for two reasons:

1. Fixed daily cost. On a construction site, there are thousands of people working and the utilization of expensive equipment (e.g. cranes). Consequently, each working day has high fixed costs.

2. The postponing of cash in-flow. Due to the postponing of cash in-flow, there are two negative effects. First, each year of construction delays the the time when cash is expect to flow into the utility increasing the interest to be paid on the debt. Second, the present value of future cash flow decreases exponentially with time.

### 2.3.1.2 Co-siting economies

Small plants allow the investors to make incremental capacity additions on a pre-existing site. This leads to co-siting economies: the set-up activities related to siting (e.g. acquisition of land rights, connection to the transmission network) have been already carried out; certain fixed indivisible costs can be saved when installing the second and subsequent units. The larger the number of NPP co-sited units, the smaller the total investment costs for each unit (Carelli et al. 2008; Carelli et al. 2010; Carelli, Petrovic & Mycoff 2007). Operational costs would also be reduced because of sharing of personnel and spare parts across multiple units (Carelli, Petrovic, Mycoff, et al. 2007) or the possibility to share the cost of upgrades on multiple units, e.g. the cost of upgrading software.

In the literature, there are many statements about co-siting economies. For example (IAEA 2005) suggests that *"the average cost for identical units on the same site can be about 15% lower than the cost of a single unit, with savings coming mostly in siting and licensing costs, site labour and common facilities. The 58 PWR in France built as multiple units at 19 sites are good examples."* A similar more recent example is the new 4-unit NPP at the Barakah site in the UAE: Unit 4 is costing some 20-25% less than Unit 1; schedule is the same by choice because the owner and operator wishes to put the units in operation at a rate of one unit per year (PC 2017). There are a number of significant benefits that could also accrue to a cluster of co-sited small NPP: consistent site output as NPPs are sequentially shut down for maintenance and refuelling; high utilization of expensive remote handling equipment; smaller components for transport and storage, meaning also more efficient use of hot cells; greater availability of spare parts and components; etc.

### 2.3.2 Modularisation and off-site fabrication

A number of papers and reports explain the costs and benefits of modularization. Most of these references are qualitative, like the recent review of modularization in the nuclear industry (Upadhyay & Jain 2016). A detailed assessment developed in the product industry, but adaptable to industrial plants is presented in (Ripperda & Krause 2017). In general modularization is the process of converting the design and construction of a monolithic plant into a plant that facilitates factory fabrication of modules for shipment and installation in the field as complete assemblies (GIF/EMWG 2007). Factory fabrication is usually cheaper than site fabrication, but the costs associated with shipping of modules to the site must also be considered. Smaller plants can take a better advantage of modularity since it is possible to have a greater percentage of factory-made components.

In the nuclear sector, (Hayns & Shepherd 1991; Schock et al. 2001) were the first authors to illustrate why the technical solutions that are embodied by the small plant design might reduce the investment cost, for a given plant. The most relevant elements of the small plant concept are the standardization of components and a broader safety by design approach. Standardization is at the origin of a more efficient supply, construction and operation (see (Langlois 2002) for a general discussion of the effects of standardization through design modularity) and it enables suppliers and utilities to more rapidly benefit the learning economies (David & Rothwell 1996). Although there are a number of works in the literature describing the qualitative advantages of modularization and prefabrication, only a few of them are able to quantify the underlying economic advantage.

A key document is (De La Torre 1994) that reviewed the work done at that time. However, the basis for the analyses are not always transparent and seems to rely more on the perception of the experts involved than on actual facts.

About construction costs:

- (Tatum 1987) state that it has been estimated that the modular engineering concept can save up to 10% of the total cost of a facility, cut on site labour by 25 %, and reduce the plot [working] area by 10% to 50 %;
- (Hesler 1990) states that *"in-depth studies have shown that modular power plants show capital cost savings of 20 % or more and schedule savings approaching 40 %"*.
- (Parkingson & Short 1982) shows examples of reduced costs through modular construction and that John Brown of John Brown Engineers & Constructors, Inc. stated that savings of at least 7% of the total contract amount was obtained by using modular construction methods rather than conventional methods for over 40 % of the process facilities for the Sullom Voe Oil Terminal in the Shetland Islands;
- (Shelley 1990) shows that in some cases, a reduction of capital costs by up to 20 % is possible. Most modular construction experts would agree that modular construction can save between 5 % and 10% of the total cost for most projects.

If construction cost can be reduced, the engineering, management and transportation costs might increase. (Glaser et al. 1979) state that the additional man hours required for design and engineering of a modular construction project increase the design and engineering cost by approximately 10%. Because of the effort needed to evaluate and select vendors, fabricators, and fabrication shops, and to administer contracts, the cost associated with procurement increases by 20% in modular construction projects, while the costs of the fabrication and transportation activities increase by approximately 17% and 13 %, respectively.

Similarly (Kliewer 1983) cited an engineering cost increase of 15 %, while (Shelley 1990) states that the transportation cost is about 1-2 % of the value of the module. (Hesler 1990) explains that the engineering costs involved in the first modular construction project are usually greater because of inexperience. In particular, the first modular design *"can be 50-60% more than conventional construction design, particularly if the job is done well. This, of course, is only 50-60% more (than conventional construction design) or 12% of the total installed cost."* However, (De La Torre 1994) concludes that despite the increased cost listed above, most modular construction projects show savings in installed costs over conventional construction.

A detailed academic study is presented by (Lapp & Golay 1997). The authors developed an analytical method, and the results of the analyses indicate a typical potential savings of 15% in the capital cost of the modular NPP versus a conventional one.

Thus, in summary, from relatively old work it seems that a target saving of 15%-20% in total installed cost is reasonable, even if there are higher up-front costs associated with engineering.

More recently (Eftimie 2016) focused on offshore facilities. According to the data owned by the author *"Standardization allows re-using same design across many different projects in a scalable, interchangeable and safe way, reducing direct project time and costs (E&C) by 20% or more, […] Reduced schedule (up to 25-50%): yard fabrication allows early procurement of critical equipment and maximized parallel works (workshop vs field civil work/site preparation); yard work can start before obtaining a site permit. Short schedules are important when required to market products rapidly […] The engineering module concept has both positive and negative aspects. Among the negative issues the most important is the cost of first design, which can exceed that of the conventional design up to 50-60% more (or reaching up to 12% of the [Total investment cost]) depending on familiarity and modular past experience of the contractor."*

A specific case study is the (General Dynamics Electric Boat 2017). Electric Boat is the prime contractor and lead design yard for the U.S. Navy's Virginia-class attack submarines. This submarine uses a 30 MW NPP. According to the company *"Improvements in construction performance will reduce construction span from 84 months to 60 months. This is being achieved through greater use of modular construction, pushing as much work as possible into a manufacturing setting where it can be done more efficiently. "*

In a recent paper (Maronati et al. 2017) developed a methodological approach (called EVAL) that using data such as the plant design and construction time, provide a target cost saving with the modularization. The authors applied the module to the Westinghouse SMR (Small modular reactors) and the authors identified possible savings for total capital investment cost in the region of 15% - 42%. However the authors did quite optimistic assumptions, for instance *"The investment required to design and develop a supply chain for the SMR under consideration was not considered […] The assumption that no limitation is present in the offsite factories production capabilities was made. Factories were assumed to be capable of producing the exact number of modules needed on site at any time, i.e. there is no constraint on the degree of parallelism allowed in the fabrication stage. Inputs for strategy two are the same as those of strategy one."* Economy of multiples and Modularisation might contribute to balance the diseconomy of scale in SMR (Trianni et al. 2009).

Despite the optimism in these papers recent experiences with the AP1000 showed as the "modules supply chain" can be a major source for delays, see for instance the case of the reactor coolant pumps (WNN 2015) and in general the issues of managing the quality along the supply chain (Hals & Flitter 2017). However these are the first AP1000 units, and perhaps the cost saving would become evident in the following units while the design and project delivery chain develop toward the NOAK.

## 2.4 The role of the project delivery chain and regulations

Nuclear experts and empirical data show that the Korean nuclear programme is probably one of the most successful. Indeed the standard Korean 1 GWe NPP is consistently built in South Korea within 5-6 years, and the fleet has one of the highest capacity factors in the world. The Koreans established a "project delivery chain" to deliver a "nuclear programme" (i.e. several, almost identical, NPPs) rather than individually commissioned NPP. Most of the time, the architect/engineer and the subcontractors were able to deliver the NPPs on time and on budget. In large projects, especially in the nuclear field, a key strategy to achieve good performances appears to be the standardization of the project delivery supply chain and NPP design. In addition to standardization (Choi et al. 2009) summarize a number of lessons to explain the success of the Korean experience:

- integration of extensive knowledge and experiences;
- strong national commitment to the NPP programme;
- continuous investment in the infrastructure with government leadership;
- localization through technology transfer (as discussed above);
- clear definition of responsibilities and rights in the NPP construction.

A clear contribution to the success of a nuclear programme is played by the regulator. Sadly famous is the role that regulation played in the USA when most of the plants were built during the 60s-80s. (Hultman et al. 2007) shows that the changing regulatory environment was a determinant for cost increases. More explicitly *"the case of nuclear power has been seen largely as an exception that reflects* *the idiosyncrasies of the regulatory environment as public opposition grew, regulations were tightened, and construction times increased […]The Gen-IV process hopes to avoid cost overruns by integrating standardized reactor designs with tighter regulatory approval timelines. It remains to be seen whether this goal can be achieved without the construction of many reactors of each type. After the accident at [Three Mile Island] in 1979, the industry was subjected to intense regulatory scrutiny and evaluation. As a result, the overall fleet capacity factor—the net generation for all reactors in the set divided by the maximum possible generation of all reactors in the set—dropped precipitously and reached its nadir in 1982 at 52.9%."*

Different from any other industrial activity, nuclear power is characterized by a strong governmental involvement due to the need for an opportune infrastructure (IAEA 2007) for both the acquisition or selling of nuclear facilities. Indeed, most of the principles governing the licensing processes are internationally recognized (IAEA 2002), but the application of these is deeply influenced by the national legal traditions and by nuclear program strategies underpinned at the country level. This differentiation is one of the most influencing factors that affect both the environment complexity and the risk from the NPP vendors and the investors' point of view. The lack of an international harmonization of "rules" is one of the greatest obstacles to the development of standardization in the nuclear industry worldwide. The risk derived from the licensing delays or no acceptance (which most of the time implies the modification of the NPP design) is especially high because the financial nature of the NPPs, which are capital intensive projects (AREVA 2006). In the 40-80's, the safety framework was mostly developed at the country level. As a result, few nuclear practices were in common between countries. After the Chernobyl accident a fundamental change of thinking was made, by governments and citizens, because the detrimental trans-boundary nature of the nuclear incidents was recognized (IAEA 2002); so an international safety framework was considered. Only after some years, the first Convention on Nuclear Safety (CNS I) was developed and implemented, and now it is adopted by almost all nuclear countries (IAEA 2010); the implementation of the CNS I was a fundamental milestone because for the first time the licensing functions where internationally recognized (article 7.2) and a general safety framework was established; the technical provisions of the CNS I are: legislation and regulation (Art: 7-9), general safety consideration (Art: 10-16) and safety of the installation (Art: 17-19).

However, in spite of agreement on these general principles, the licensing framework is still different between the nations.

An analysis of the licensing challenge to deploy SMR is presented in (Sainati et al. 2015). Key points are:

- The type of licensing approach is a fundamental determinant for the deployment of SMRs. At this stage of development, the "goal setting approach" seems the most favorable to the deployment for SMR. Conversely, most of the countries involved (as NPP vendor, buyer or both) in an SMR nuclear program adopt a prescriptive based licensing approach.
- The existing licensing processes have been developed having implicitly in mind GWe scale NPPs that have multi-billions budget and 5-10 years construction. Therefore the licensing process could extend the construction time of SMRs beyond the pure technical schedule undermining the overall economics.
- As aforementioned, the fragmentation at country level of legal systems and jurisprudence, institutional systems, licensing process structure all constrain the NPP standardization. Since each country has jurisdiction only on itself, a short-term harmonization is unlikely.
- The idea of "NPP certified in the factory" is still far away from practice. Even if all the "mechanical components" of an NPP are certified in the factory, the licensing process applies to another unit of analysis: the system installed at the site. The nuclear operator is in all cases the ultimate and sole responsible party for nuclear safety.
- An ad-hoc legislation process similar to the one for research reactors could be the way forward. However, there could be constraints in terms of public acceptability, and total power capacity installed at the site. Research reactors are designed as stand-alone systems, not duplicated at site and (usually) produce limited or no power.

In conclusion, the project delivery chain is a key determinant of the success of megaprojects. In particular, for the nuclear sector, standardisation and regulation/licensing are the foremost topic that needs to be considered.

# 3 Conclusions

Megaprojects are often delivered overbudget and late. Their size is a key driver, being directly correlated with the likelihood and magnitude of their overbudget and delay. NPPs are not an exception to this rule, and currently, all NPPs under construction in Europe and USA are delivered overbudget and late. However, the nuclear technology is not to blame, for example, in South Korea NPPs are delivered on time & within the budget and have very good operational performance. The reasons for this success can be found in the project delivery chain. NPPs or any other type of megaprojects are likely to be delivered overbudget and late when the project is a complex "one-of-its-kind white elephant" combining a novel complex technology with a complex network of stakeholders. This combination of novelty and complexity make the megaproject fragile, i.e. it is likely that something will go wrong and the project will be very difficult to recover.

The proven way to address this issue is *standardisation*, which played a key role in the South Korea success. Standardization needs to be twofold: the technical standardisation, i.e. the construction of the infrastructure with the same (or very similar) design over and over, and the "project delivery chain" standardisation, i.e. the same stakeholders involved in the delivery of a project that is replicable multiple times. Under this perspective (SMRs), given their size, are an ideal power plant for several countries. Yet, if the economy of scale is the only driver considered, SMRs are hardly competitive with large NPPs (and with gas or coal power plants). However, a fleet of standard SMRs might balance the "diseconomy of scale" with the "economy of multiples." The delivery of several standardised SMR projects might be the key to achieving good project management performances in the nuclear sector. The deployment of SMRs, however, faces a number of challenges from the licencing, supply chain and financing perspective. These challenges might be enormous, but so too are the potential rewards.

# Acknowledgments

This documents summarises what the author learned in the last 10 years of research in megaprojects and nuclear power plants. Countless discussions and research meetings with industry experts, policymakers, academic colleagues and other experts contributed to developing the background knowledge to write this document, and the author is very grateful to all these persons. Specifically for this document, the author acknowledge the valuable support from Prof Jacopo Buongiorno (MIT) and Dr. David Petti (INL) that reviewed many times the draft versions and provided very valuable feedback. The discussion with Prof. Naomi Brookes, Prof. Nigel Smith, Diletta Colette Invernizzi and Tristano Sainati have also been incredibly useful. The author remains the only person accountable for omissions and mistakes.

Quantification of Impacts to Support Decision Making. *Journal of Mechanical Design*, 139(2), p.021103.

Ruuska, I. et al., 2009. Dimensions of distance in a project network: Exploring Olkiluoto 3 nuclear power plant project. *International Journal of Project Management*, 27(2), pp.142–153.

Ryan, M.J. & Faulconbridge, R.I., 2005. *Engineering a System: Managing Complex Technical Projects*,

Sainati, T., Locatelli, G. & Brookes, N., 2015. Small Modular Reactors: Licensing constraints and the way forward. *Energy*, 82, pp.1092–1095.

Sainati, T., Locatelli, G. & Brookes, N., 2017. Special Purpose Entities in Megaprojects: empty boxes or real companies? *Project Management Journal*, 48(2).

Schock, R.N., Brown, N.W. & Smith, C.F., 2001. *Nuclear Power, Small Nuclear Technology, and the Role of Technical Innovation: An Assessment*,

Shelley, S., 1990. Making inroads with modular construction. *Chemical Engineering*, August, pp.30–35.

Sovacool, B.K., Gilbert, A. & Nugent, D., 2014a. An international comparative assessment of construction cost overruns for electricity infrastructure. *Energy Research & Social Science*, 3, pp.152–160.

Sovacool, B.K., Gilbert, A. & Nugent, D., 2014b. Risk, innovation, electricity infrastructure and construction cost overruns: Testing six hypotheses. *Energy*, 74, pp.906–917.

Sovacool, B.K., Nugent, D. & Gilbert, A., 2014. Construction Cost Overruns and Electricity Infrastructure: An Unavoidable Risk? *The Electricity Journal*, 27(4), pp.112–120.

Tatum, C.B., 1987. Improving Constructibility during Conceptual Planning. *Journal of Construction Engineering and Management*, 113(2), p.191.

Trianni, A., Locatelli, G. & Trucco, P., 2009. Competitiveness of small-medium reactors: A probabilistic study on the economy of scale factor. In *International Congress on Advances in Nuclear Power Plants 2009, ICAPP 2009*.

Turner, R. & Zolin, R., 2012. Forecasting Success on Large Projects: Developing Reliable Scales to Predict Multiple Perspectives by Multiple Stakeholders Over Multiple Time Frames. *Project Management Journal*, 43(5), pp.87–99.

Upadhyay, A.K. & Jain, K., 2016. Modularity in nuclear power plants: a review. *Journal of Engineering, Design and Technology*, 14(3), pp.526–542.

Wang, G. et al., 2017. Exploring the impact of megaproject environmental responsibility on organizational citizenship behaviors for the environment: A social identity perspective. *International Journal of Project Management*, In press.

Warrack, A.A., 1985. *Resource Megaproject Analysis and Decision Making*, Western Resources Program, Institute for Research on Public Policy.
27